# Irreducible decomposition of Gaussian distributions

# and the spectrum of black-body radiation

**Sándor Varró**

Research Institute for Solid State Physics and Optics

of the Hungarian Academy of Sciences

H-1525 Budapest, POBox 49, Hungary

e-mail : varro@sunserv.kfki.hu

**Abstract**

It is shown that the energy of a mode of a classical chaotic field, following the continuous exponential distribution as a classical random variable, can be uniquely decomposed into a sum of its fractional part and of its integer part. The integer part is a discrete random variable whose distribution is just the Bose distribution yielding the Planck's law of black-body radiation. The fractional part – as we call it – is the "dark part" with a continuous distribution, which is, of course, not observed in the experiments. It is proved that the Planck-Bose distribution is infinitely divisible, and the irreducible decomposition of it is given. The Planck variable can be decomposed into an infinite sum of independent binary random variables representing the "binary photons" – more accurately photo-molecules or multiplets – of energy $2^s h\nu$ with s = 0, 1, 2, … . These binary photons follow the Fermi statistics. Consequently the black-body radiation can be viewed as a mixture of thermodynamically independent fermion gases consisting of "binary photons". The binary photons give a natural tool for the dyadic expansion of an arbitrary (but not coherent) ordinary photon excitations. It is shown that the binary photons have wave-particle fluctuations of fermions. These fluctuations combine to give the wave-particle fluctuations of the original bosonic photons, expressed by Einstein's fluctuation formula.

1. Introduction

The concept of energy quanta has long been introduced by Boltzmann [1] in 1877, when he calculated the number of distributions of them among the molecules of an ideal gas being in thermal equilibrium. He considered these discrete indistinguishable energy elements as mere a mathematical tool in his combinatorial analysis, and was able to show that the "permutability measure" found this way can be identified with the entropy. A similar idea was used by Planck [2] in 1900 when he, besides deriving the correct spectrum of black-body radiation, discovered the new universal constant, the elementary quantum of action $h = 6.626 \times 10^{-27} erg \cdot \sec$. The study of black-body radiation has played a crucial role in the development of quantum physics, and still today it is an important element in many branches of investigations (see for instance the analysis of thermal noise in communication channels or the role of the cosmic microwave background radiation played in cosmological physics). One would think that not much new can be said about the nature of black-body radiation which has a well-established theoretical foundation in modern physics. In the present paper we hope to show that there are still some aspects of this subject which may be worth to explore. Our analysis is not conventional in the sense that it relies exclusively on classical probability theory, and does not use field quantization in the modern sense. In this respect the paper is written in the spirit of the early investigations by Planck, Einstein and others, but the basic results to be presented here are based on modern probability theory, which was born at the beginning of the thirties, so it had not been at disposal at the time of Planck's original investigations. In order to put our results into a proper perspective, we think, it is worth to give a brief overview of the basic facts and ideas concerning the theory of black-body radiation.

Planck expressed the spectral energy density (the fraction of energy per unit volume in the spectral range $(\nu, \nu + d\nu)$) of the black-body radiation as $u_\nu = (8\pi\nu^2/c^3)U_\nu$, where $U_\nu$ denotes the average energy of one Hertzian oscillator being in thermal equilibrium with the radiation. Let us remark that the first factor $Z_\nu = 8\pi\nu^2/c^3$, which came out from the dynamics of the Hertzian oscillator in Planck's analysis, is, on the other hand, nothing else but the spectral mode density of the radiation field enclosed in a *Hohlraum*, in a cavity bounded by perfectly reflecting walls. So $VZ_\nu d\nu$ is the number of modes in the volume $V$ in the spectral range $(\nu, \nu + d\nu)$. Planck calculated the entropy of the oscillator system by using combinatorial analysis, where he wrote down the total energy of the oscillators as an integer multiple of an energy element $\varepsilon_0$. Then, by taking Wien's displacement law into account (according to which $u_\nu$ must be of the form $u_\nu = \nu^3 F(\nu/T)$, where $F$ is an universal function), he concluded that the energy element must be proportional to the frequency of the spectral component under discussion, i.e. $\varepsilon_0 = h\nu$. The factor of proportionality is now called Planck's constant. The well-known Planck formula for the spectral energy density so obtained reads

$$u(\nu,T) = Z_\nu U_\nu = Z_\nu \frac{\varepsilon_0}{e^{\varepsilon_0/kT}-1} = \frac{8\pi\nu^2}{c^3} \frac{h\nu}{e^{h\nu/kT}-1},$$

where $c$ denotes the velocity of light *in vacuo*, $k = 1.831 \times 10^{-16} erg/K$ is the Boltzmann constant and $T$ is the absolut temperature. This formula was in complete accord with the precise experimental results by Pringsheim and Lummer [3], in the short wavelength regime, and by Rubens and Kurlbaum [4] in the long wavelength regime. It is remarkable that, on the basis of the four universal constants $c$, $k$, $h$ and $G$ (the latter being Newton's gravitational constant), as Planck realized, it is possible to introduce the *natural system of units* of length ($l_P = \sqrt{\hbar G/c^3} = 1.616 \times 10^{-33} cm$), time ($t_P = l_P/c = 5.392 \times 10^{-44} s$), mass ($m_P = \sqrt{\hbar c/G} = 2.176 \times 10^{-5} g$) and temperature ($T_P = m_P c^2/k = 1.417 \times 10^{32} K$). We have used the modern notation $\hbar = h/2\pi$, and the nowadays available experimental values of the universal constants. Let us remark here that the black-body radiation and Planck's formula has received a renewed interest and importance since 1965, when Penzias and Wilson discovered the cosmic microwave background radiation, which is one of the main witnesses of the big bang theory of the universe [5]. The spectrum of this $2.728 \pm 0.004\ K$ black-body radiation measured in the FIRAS experiment by the COBE satellite [5], fits excellently well to Planck's formula. They say that such a perfect agreement has not been obtained so far in laboratory experiments using artificial black-bodies.

In 1905 Einstein introduced the concept of light quanta (photons) [6] on the basis of a thermodynamical analysis of the entropy of black-body radiation in the Wien limit of Planck's formula,

$$u(\nu,T) \approx \rho'' = Z_\nu U_\nu'' \equiv \frac{8\pi\nu^2}{c^3} h\nu \cdot e^{-h\nu/kT} \quad (h\nu/kT \gg 1).$$

On passing, we may safely state that all the other later results by Einstein concerning photons were exclusively based on the study of black-body radiation (see e.g. Refs. [7] and [8]). Einstein noted in 1907 [7] that the correct average energy $U_\nu$ of one oscillator obtained by Planck can be calculated from the Boltzmann distribution by using the replacement

$$U'_\nu \equiv \frac{\int_0^\infty dE\, E\, e^{-E/kT}}{\int_0^\infty dE\, e^{-E/kT}} \to U_\nu = \frac{\sum_{n=0}^\infty n\varepsilon_0 e^{-n\varepsilon_0/kT}}{\sum_{n=0}^\infty e^{-n\varepsilon_0/kT}} = \frac{\varepsilon_0}{e^{\varepsilon_0/kT}-1} = \frac{h\nu}{e^{h\nu/kT}-1}.$$

Here one has to keep in mind that the energy integral originally comes from a two-dimensional phase-space integral ($dpdq \to dE$) of the one-dimensional oscillator. (In higher dimensions the density of states is not constant, as here.) It is clear that the above replacement corrensponds to restrict the integrations in relatively very narrow ranges of the energy around the integer multiples of the energy quantum $\varepsilon_0$. Roughly speaking, we obtain the Planck factor if take the "*integer part of the Boltzmann distribution*". *Without this discretization* we obtain $U'_\nu = kT$ (which expresses the equipartition of energy), and we end up with the Rayleigh-Jeans formula,

$$u(\nu,T) \approx \rho' = Z_\nu U'_\nu = \frac{8\pi\nu^2}{c^3}kT \quad (h\nu/kT \ll 1),$$

which otherwise can also be derived from the Planck formula in the limit of large radiation densities, as is indicated after the above equation. Einstein's hypoteses on light quanta received a new support in 1909 by the famous fluctuation formula [8], which contains both particle-like and wave-like fluctuation of the energy of black-body radiation occupying a sub-volume of the *Hohlraum*. This was the first mathematically precise formulation of the wave-particle duality. After Bohr's atom model had appeared, Einstein showed in 1917 [9] that the Planck formula can be obtained from a "reaction kinetic" consideration in which one takes into account the detailed balance of spontaneous and induced emission and the absorption processes of material systems with discrete energy levels being in thermal equilibrium with the radiation.

In 1910 Debye reconsidered the problem of the spectrum of black-body radiation on a completely different basis [10], namely he left out of discussion Planck's resonators, and he directly applied the combinatorial analysis and the entropy maximization procedure to the energies of the *normal modes themselves*, and in this way he was able to derive the correct (Planck) formula. Later Wolfke [11] and Bothe [12] showed that the black-body radiation is equivalent to a mixture of infinitely many thermodynamically independent classical gases consisting of so-called "photo-molecules" (or photo-multiplets) of energies $h\nu$, $2h\nu$, $3h\nu$, and so on. This approach and some historical aspects to be outlined shortly below have been discussed in detail in our recent paper [22].

In 1911 Natanson [13] used again Boltzmann's original method to find the most probable distribution of energy quanta on energy "receptacles" (which may mean either Planck's resonators (or other ponderable material particles) or cavity normal modes (phase-space cells)) and derived the correct equilibrium (Bose) distribution. The same method was rediscovered by Bose 13 years later in his famous derivation of Planck's law of black-body radiation [14]. The new statistics was first applied to an ideal gas by Einstein [15], and the existence of Bose-Einstein condensation was predicted in [16]. It is interesting to note that Natanson's general formulae obtained by Boltzmann's original method, can directly be applied to particles whose occupation numbers are restricted by the Pauli exlusion principle. From the point of view of real physical consequences, this case was first studied by Fermi [17] in 1926. An alternative "reaction kinetic" derivation of the Fermi distribution was given one year later by Ornstein and Kramers [18].

Concerning the general theory of black-body radiation we refer the reader to the classic book by Planck [19]. For further details on the development of the black-body theory and the concept of energy quanta, see Ref. [20], and, in particular, the very thoroughly written book by Kuhn [21]. In our recent work [22] a historical overview concerning Einstein's fluctuation formula can be found, and some results to be presented below have also been briefly outlined there in a different context.

In 1932 Schweikert realized [23] that the Planck factor can be derived from the Boltzmann distribution *without any appearent discretization*,

$$U_\nu = U'_\nu - U'''_\nu \equiv \frac{\int_0^\infty dE E e^{-E/kT}}{\int_0^\infty dE e^{-E/kT}} - \frac{\int_0^{\varepsilon_0} dE E e^{-E/kT}}{\int_0^{\varepsilon_0} dE e^{-E/kT}} = kT - \left(kT - \frac{\varepsilon_0}{e^{\varepsilon_0/kT}-1}\right) = \frac{h\nu}{e^{h\nu/kT}-1}.$$

According to Schweikert, the above formula corresponds to the physical picture, that only those oscillators (atoms) contribute to the thermal radiation (at a particular frequency $\nu$) whose energies are above the threshold value $\varepsilon_0 = h\nu$. On the other hand, Schweikert argued that, on the ground of classical statistics, it cannot be seen why should one use a *difference of two mean values* (namely $U'_\nu$ and $U'''_\nu$), instead, why not to *directly subtract the contributions* coming from the lower energies ($0 < E < \varepsilon_0$), and use the original normalization factor, i.e.

$$\tilde{U}_\nu \equiv \frac{\int_0^\infty dE E e^{-E/kT} - \int_0^{\varepsilon_0} dE E e^{-E/kT}}{\int_0^\infty dE e^{-E/kT}} = (kT + h\nu)e^{-h\nu/kT}.$$

The spectral energy density $\tilde{u}(\nu,T) = Z_\nu \tilde{U}_\nu$ obtained on the basis of the above "average" energy $\tilde{U}_\nu$ interpolates between the Rayleigh-Jeans and the Wien formulae, but, needless to say, it is not that accurate, as the exact Planck formula. Nevertheless, Schweikert favoured his new radiation formula to that of Planck, because the former one was derived from the continuous Boltzmann distribution and no additional assumption on the quantization of energy was needed. He argued that further and more accurate experimental data were needed to check which of the two formulae corresponds better to reality. After all, in the mid-thirties of the last century, concerning direct measurements of the black-body spectrum, this view could not have been thought to be completely unjustified.

The moral for us from Schweikert's analysis is that if one substracts from the average energy ($U'_\nu$), coming from the complete Boltzmann distribution, the average ($U'''_\nu$) coming from the range of the *fraction* of $\varepsilon_0$, i.e. from the interval $0 < E < \varepsilon_0$, then one receives the correct Planck factor. After all, this conclusion is in accord with Einstein's observation, according to which the integer part of the energy determines the spectrum. In the forthcoming sections we will show that there can be much more said about the division of the Boltzmann distribution, namely, not only at the level of expectation values, but at the level of the distribution itself. We will show that it is possible to divide this distribution into the product of the distributions of its (irreducible) fractional part and of its integer part, the latter being the Planck-Bose distribution. The Planck-Bose distribution *deduced* in such a way can be further decomposed into a product of irreducible distributions corresponding to binary variables ("binary photons", as we propose to call them), which have fermionic character. In this way, the original Gauss random variables characterizing the field amplitudes of the black-body radiation (from which we derive the Gauss variable) at a given frequency, are uniqely decomposed into a sum of irreducible variables which cannot be divided any further. We emphasize that throughout the paper we do not need the *explicite* use of Planck's resonators or any other material agents being in thermal equilibrium with the radiation. We merely assume that the radiation consist of independent modes of spectral mode density $Z_\nu$ given above, and we consider one of these modes whose (two independent) amplitudes (according to the concept of "molecular disorder" and to the central limit theorem) are assumed to follow the (completely chaotic) Gauss distribution.

In section 2 we derive the Rayleigh-Jeans formula from the central limit theorem of classical probability theory. This section also serves as an introduction of our basic notations and the philosophy of the present paper. In Section 3 and 4 it is shown that the energy of a mode of a classical chaotic field, following the continuous exponential distribution as a classical random variable, can be uniquely decomposed into a sum of its fractional part and of its integer part. The integer part is a discrete random variable whose distribution is just the Planck-Bose distribution yielding the Planck's law of black-body radiation. The fractional part – as we call it – is the "dark part" with a continuous distribution, is, of course, not observed in the experiments. In Section 5 it is proved that the Planck-Bose distribution is infinitely divisible. The Planck variable can be decomposed into an infinite sum of independent binary random variables representing the "binary photons" – more accurately photo-molecules or multiplets – of energy $2^s h\nu$ with s = 0, 1, 2, … . These binary photons follow the Fermi statistics. Consequently the black-body radiation can be viewed as a mixture of thermodynamically independent fermion gases consisting of binary photons. In section 6 a short summary closes our paper.

## 2. Derivation of the Rayleigh-Jeans formula from the central limit theorem

In classical physics the black-body radiation, a radiation being in thermal equilibrium in a *Hohlraum* (a cavity with perfectly reflecting walls at absolute temperature *T*) is considered as a chaotic electromagnetic radiation. The average spatial distribution of such a stationary radiation is homogeneous and isotropic and the electric field strength and the magnetic induction of its spectral components have completely random amplitudes which are built up of infinitely many independent infinitesimal contributions. In this description the electric field strength and the magnetic induction of a mode (characterized by its frequency $\nu$, wave vector and polarization) of the thermal radiation (in a small spatial region) are proportional with the random process

$$a_\nu(t) = a_c \cos(2\pi\nu t) + a_s \sin(2\pi\nu t) = \sqrt{a_c^2 + a_s^2} \cos[(2\pi\nu t) - \theta], \quad \theta = \arg(a_c + ia_s) \quad (1)$$

where $a_c$ and $a_s$ are independent random variables. According to the *central limit theorem* of classical probability theory – under quite general conditions satisfied by the otherwise *arbitrary* distributions of the mentioned infinitesimal amplitude elements – the asympthotic *probability distributions* of the resultant amplitudes necessarily approach *Gaussian distributions* expressed by the error function $\Phi(x)$. The first precise formulation of this theorem is due to Lindenberg [27]. For our purposes here a theorem on the limit behaviour of *probability density functions* due to Gnedenko suits better (see e. g. Ref. [27], p.370). Let $a_1,...,a_k,...,a_n$ and $a'_1,...,a'_k,...,a'_n$ be completely independent random variables of the same probability density function $f$ with zero expectation values and of a common finite variance $a^2$. Then the probability density functions $f_n$ of the normalized superpositions

$$a_{cn}/a = (a_1 + ... + a_k + ... + a_n)/(a\sqrt{n}), \quad a_{sn}/a = (a'_1 + ... + a'_k + ... + a'_n)/(a\sqrt{n}) \qquad (2)$$

go over to Gaussian probability densities in the limit $n \to \infty$,

$$P(x \leq a_{cn}/a < x + dx) = f_n(x)dx \to (2\pi)^{-1/2} \exp(-x^2/2)dx, \qquad (3)$$

and a similar relation for the sine component $a_{sn}/a$. Hence the amplitudes $a_c$ and $a_s$ in Eq. (1) may be considered as independent Gaussian random variables, i.e.

$$P(q \leq a_c < q + dq, \ p \leq a_s < p + dp) = P(q \leq a_c < q + dq)P(p \leq a_s < p + dp)$$
$$= [f_c(q)dq][f_s(p)dp] = \left[\frac{1}{a\sqrt{2\pi}}\exp\left(-q^2/2a^2\right)dq\right]\left[\frac{1}{a\sqrt{2\pi}}\exp\left(-p^2/2a^2\right)dp\right]. \qquad (4)$$

The physical meaning of the parameter $a$ can be obtained by requireing that the average spectral energy density $u_\nu$ be equal to the product of the spectral mode density $Z_\nu = 8\pi\nu^2/c^3$ and the average energy $\bar{\varepsilon}$ of one mode, i.e. $u_\nu = \overline{a_\nu^2(t)}/8\pi = a^2/8\pi = Z_\nu\bar{\varepsilon}$, where $u_\nu d\nu$ gives the energy of the chaotic radiation per unit volume in the spectral range $(\nu, \nu + d\nu)$. By introducing the mode energy $E$ as a classical random variable by the definition $(a_c^2 + a_s^2)/16\pi = Z_\nu E$, the joint probability given by Eq. (4) can be expressed in terms of the new "action-angle variables" ($\varepsilon = (q^2 + p^2)/Z_\nu 16\pi$ and $\vartheta = \arg(q + ip)$) :

$$P(q \leq a_c < q + dq, \ p \leq a_s < p + dp)$$
$$= P(\varepsilon \leq E < \varepsilon + d\varepsilon)P(\vartheta \leq \theta < \vartheta + d\vartheta) = \left[(1/\bar{\varepsilon})\exp(-\varepsilon/\bar{\varepsilon})d\varepsilon\right](d\vartheta/2\pi) \cdot \quad (5)$$

According to Eq. (5) the energy of each mode of the chaotic field is an exponential (Boltzmann) random variable, and the phases are distributed uniformly.

From the point of view of our analysis, it is crutial to introduce *two* independent energy parameters $\varepsilon_0$ and $\bar{\varepsilon}$ (containing *two different universal constants*, namely the Planck constant and the Boltzmann constant) with the help of which we define the dimensionless energy variables and their probability density distributions. First we introduce the scaled energy $\eta \equiv E/\varepsilon_0$ of a mode of the chaotic field and the parameter $\beta = \varepsilon_0/\bar{\varepsilon}$, and henceforth we shall call $\eta$ as *Gauss variable* (because, though it satisfies the (two-dimensional) Boltzmann distribution, it stems originally from the Gaussian chaotic amplitudes). By taking Eq. (5) into account the dimensionless probability density function $f_\eta(y)$ of $\eta$ and its expectation value are given by the relations

$$P(y \leq \eta < y + dy) = f_\eta(y)dy, \ f_\eta(y) = \beta e^{-\beta y}, \ (0 \leq y < \infty), \quad (6)$$

$$\bar{\eta} \equiv \int_0^\infty dy f_\eta(y) y = \frac{1}{\beta} \equiv \frac{1}{(\varepsilon_0/\bar{\varepsilon})} \rightarrow E_\eta \equiv \varepsilon_0 \bar{\eta} = \frac{\varepsilon_0}{(\varepsilon_0/\bar{\varepsilon})} = \bar{\varepsilon} \ . \quad (7)$$

According to Boltzmann's principle, the entropy $S_\eta$ of a chaotic mode is given as

$$S_\eta(E_\eta) \equiv -k\int_0^\infty f_\eta(y)\log f_\eta(y)dy = k(1-\log\beta) = k\log(e\bar{\eta}) = k\log(eE_\eta/\varepsilon_0), \quad E_\eta = \bar{\varepsilon}, \quad (8)$$

where $k = 1.381 \times 10^{-16} erg/K$ denotes the Boltzmann constant. From the fundamental relation $\partial S/\partial E = 1/T$ of phenomenological thermodynamics (by taking into account that $\partial S_\eta/\partial\bar{\eta} = k\beta$) we obtain

$$\partial S_\eta/\partial E_\eta = 1/T \ \rightarrow \ E_\eta = \bar{\varepsilon} = kT \ , \ \partial^2 S_\eta/\partial E_\eta^2 = -k/E_\eta^2 \ . \quad (9)$$

In this way we have got closer to the physical meaning of the parameter $\beta$, namely we have $\beta = \varepsilon_0 / kT$. The second equation $E_\eta = \overline{\varepsilon} = kT$ in Eq. (9) expresses the *equipartition of energy*, which means in the present case that the average energy of the modes are the same, regardless of their frequencies, propagation direction and polarization. We may say that $kT/2$ energy falls on average to each quadratic term of the radiant energy of each mode. If we multiply the average energy $kT$ of one mode with the spectral mode density $Z_\nu$ then we obtain the spectral energy density $u_\nu = \rho' = u^{R-J}(\nu,T) = (8\pi\nu^2/c^3)kT$, which is called the *Rayleigh-Jeans formula*. It describes quite well the experimental results for low frequencies, but for large frequencies it diverges (this artefact has been termed as "ultraviolet catastrophe"). From Eq. (6) the variance of $\eta$ is simply determined,

$$\Delta\eta^2 \equiv \int_0^\infty dy f_\eta(y)(y-\overline{\eta})^2 = \overline{\eta^2} - \overline{\eta}^2 = \frac{1}{\beta^2} = \overline{\eta}^2. \tag{10}$$

In a sub-volume $\upsilon$ of the *Hohlraum* in the spectral range $(\nu, \nu + d\nu)$ the number of modes $m_\nu$ and the total energy of them are given, respectively, as

$$m_\nu = \upsilon Z_\nu d\nu = \upsilon \frac{8\pi\nu^2 d\nu}{c^3}, \quad \overline{E}_\nu = m_\nu E_\eta = m_\nu \varepsilon_0 \overline{\eta} = m_\nu kT, \tag{11}$$

hence the fluctuation (variance) of the energy can be brought to the form

$$\Delta E_\nu^2 = m_\nu \varepsilon_0^2 \Delta\eta^2 = \frac{\overline{E}_\nu^2}{m_\nu} = \frac{c^3}{8\pi\nu^2 d\nu} \frac{\overline{E}_\nu^2}{\upsilon}. \tag{12}$$

The expression on the right-hand-side of Eq. (12) is *formally* equivalent to the so-called *wave-like fluctuation* of the energy of the black-body radiation in Einstein's famous fluctuation formula [8]. Notice that in all the physical results expressed by Eqs. (9), (11) and (12) the energy scaling parameter $\varepsilon_0$ does not show up at all, it drops out from all the final formulae.

So, in all the above results only *one* universal parameter is present, namely the Boltzmann constant $k$.

**3. The fractional part of the Gauss variable**

The fractional part $z = \{y\} \equiv y - [y] \equiv \psi(y)$ of a real variable $0 \leq y < \infty$ is a strictly increasing function on the intervals $(k, k+1)$, where $k = 0, 1, 2, \ldots$, so its inverse function $y = \psi^{-1}(z) = y_k(z) = z + k$ $(k = 0, 1, 2, \ldots)$ is piece-wise continuously differentiable on this countable set of intervals. Hence the probability density function of the fractional part $\varsigma \equiv \{\eta\}$ of the random variable $\eta$ of the exponential distribution $f_\eta(y)$, Eq. (6), can be calculated by using the general formula (see e. g. Ref. [27], p. 161)

$$f_\varsigma(z) = \sum_{k=0}^{\infty} \frac{f_\eta(\psi_k^{-1}(z))}{|\psi_k'(\psi_k^{-1})|} = \sum_{k=0}^{\infty} f_\eta(z+k)\left|\frac{d(z+k)}{dz}\right| = \beta e^{-\beta z}\sum_{k=0}^{\infty} e^{-k\beta},$$

$$\varsigma \equiv \{\eta\}, \quad P(z \leq \varsigma < z + dz) = f_\varsigma(z)dz, \quad f_\varsigma(z) = \beta\, e^{-\beta z}/(1-e^{-\beta}), \quad (0 \leq z < 1). \tag{14}$$

The expectation value $\bar{\varsigma}$ determines the average energy of the fractional part,

$$\bar{\varsigma} = \int_0^1 dz\, z f_\varsigma(z) = \frac{1}{\beta} - \frac{1}{e^\beta - 1} \to E_\varsigma \equiv \varepsilon_0 \bar{\varsigma} = \varepsilon_0\left[\frac{1}{(\varepsilon_0/kT)} - \frac{1}{\exp(\varepsilon_0/kT)-1}\right]. \tag{15}$$

Notice that the by now unknown energy scale parameter does not drop out. From Eqs. (7), (9) and (15) we receive the Planck factor if we substract the expectation value of the fractional part from the expectation value of the Gauss energy

$$E_\eta - E_\varsigma = \varepsilon_0(\bar{\eta} - \bar{\varsigma}) = \frac{\varepsilon_0}{e^{\varepsilon_0/kT}-1}. \tag{16}$$

By multiplying with the spectral mode density $Z_\nu$ we obtain Planck's formula.

$$u(\nu,T) = Z_\nu(E_\eta - E_\varsigma) = \frac{8\pi\nu^2}{c^3}\frac{\varepsilon_0}{e^{\varepsilon_0/kT}-1} = \frac{8\pi\nu^2}{c^3}\frac{h\nu}{e^{h\nu/kT}-1}. \tag{17}$$

The last equation of Eq. (17) comes from Wien's displacement law, which states that the spectral energy density has to be of the form $u(\nu, T) = \nu^3 F(\nu/T)$, with $F$ being a universal function. Accordingly, the energy parameter $\varepsilon_0 = h\nu$ has to be proportional with the frequency, and the factor of proportionality has to be chosen Planck's elementary quantum of action $h = 6.626 \times 10^{-27} erg.\sec$ (if one wishes to get agreement with the experimental data). Since Eq. (17) coincides with the measured spectrum of thermal radiation with an unprecedented accuracy, it is clear that the contribution of the fractional part $\varsigma$ is not measured, so it is justified to call this part the "*dark part*" of the energy of the chaotic field. Henceforth we shall call $\varsigma$ the *dark variable*. It is remarkable that, according to Eq. (15), for high temperatures or/and for small frequencies (more accurately, in the limit $\beta = (h\nu/kT) \to 0$) the energy of the "dark part" approaches *from below* the zero-point energy, i. e. $E_\varsigma \to h\nu/2$:

$$\lim_{\beta \to 0} \bar{\varsigma} = \int_0^1 dz\, z \lim_{\beta \to 0} f_\varsigma(z) = \int_0^1 dz\, z = \frac{1}{2}, \ i.e.\ \lim E_\varsigma = \frac{h\nu}{2} \ (h\nu/kT \to 0), \quad (18)$$

and for zero temperature it goes to zero ($E_\varsigma \to 0$ for $T \to 0$). The entropy of the dark part reads:

$$\begin{aligned} S_\varsigma &= -k \int_0^1 dz f_\varsigma(z) \log f_\varsigma(z) = k(1 - \log \beta) - k\left[\beta\left(1 + \frac{1}{e^\beta - 1}\right) + \log\left(\frac{1}{e^\beta - 1}\right)\right] \\ &= k \log(e\bar{\eta}) - k[(1 + \bar{\eta} - \bar{\varsigma})\log(1 + \bar{\eta} - \bar{\varsigma}) - (\bar{\eta} - \bar{\varsigma})\log(\bar{\eta} - \bar{\varsigma})] \\ &= S_\eta - k\{[1 + (E_\eta - E_\varsigma)/h\nu]\log[1 + (E_\eta - E_\varsigma)/h\nu] - [(E_\eta - E_\varsigma)/h\nu]\log[(E_\eta - E_\varsigma)/h\nu]\} \end{aligned}, \quad (19)$$

$$S_\varsigma = S_\eta - k[(1 + \bar{n})\log(1 + \bar{n}) - \bar{n}\log\bar{n}], \quad \bar{n} \equiv \bar{\eta} - \bar{\varsigma} = \frac{1}{e^{h\nu/kT} - 1}. \quad (20)$$

From the usual relation $\partial S/\partial E = 1/T$ of thermodynamics (by taking into account that $\partial S_\varsigma/\partial\bar{\varsigma} = k\beta = k\varepsilon_0/kT$) we have

$$\partial S_\varsigma / \partial E_\varsigma = \partial S_\eta / \partial E_\eta = 1/T , \tag{21}$$

which means that the subsystem characterized by $\varsigma$ has the same temperature as $\eta$. The fluctuation of the "dark variable" contains both the bosonic particle-like term with "particle energy" $(2\bar{n}-1)h\nu$ and a wave term which has the same form as in Einstein's fluctuation formula,

$$\Delta E_\varsigma^2 = (2\bar{n}-1)h\nu\overline{E_\varsigma} + \overline{E_\varsigma}^2 / M_\nu . \tag{21a}$$

Next we prove, that from our formalism not only the correct *average energy*, Eq. (16), but also the correct *energy distribution* (photon number distribution) of the radiation comes out.

**4. The Planck-Bose part of the chaotic field**

It is known that the exponential distribution (6) – as a special case of the gamma distribution – is an infinitely divisible distribution. A random variable $\eta$ is said to be infinitely divisible if for any natural number $n$, it can be decomposed into a sum of completely independent random variables having the same distribution: $\eta = \eta_1 + \eta_2 + ... + \eta_n$ [27]. In this case the characteristic function (which is the Fourier transform of the probability density distribution) is also infinitely divisible [26], which means that the characteristic function of the sum is a product of the characteristic functions of the summands. The characteristic function of the exponential distribution reads

$$\varphi_\eta(t) \equiv \langle e^{i\eta \cdot t} \rangle = \int_0^\infty dy f_\eta(y) e^{iy \cdot t} = \left(1 - \frac{it}{\beta}\right)^{-1} , \tag{22}$$

where the bracket denotes expectation value. It is clear that $\varphi_\eta(t) = [\varphi_n(t)]^n$, where $\varphi_n(t) = (1-it/\beta)^{-1/n}$ are characteristic functions of the same gamma distributions with the parameter $1/n$ which is the common distribution of the variables $\eta_1, \eta_2, ..., \eta_n$ :

$f_n(y) = [\beta^{1/n}/\Gamma(1/n)] y^{(1/n)-1} \exp(-\beta \cdot y)$. However there exist another decomposition of the exponential distribution which we may already be suspected from the above analysis of its fractional part, Eq. (13), namely, $\eta$ can be uniquely decomposed into a sum of its fractional part and of its integer part. This can be best viewed by calculating first the characteristic function of $\varsigma = \{\eta\}$, which, according to Eq. (14) reads

$$\varphi_\varsigma(t) \equiv \langle e^{i\varsigma \cdot t} \rangle = \int_0^1 dz f_\varsigma(z) e^{iz \cdot t} = \left(1 - \frac{it}{\beta}\right)^{-1} \left(\frac{1 - be^{it}}{1 - b}\right), \quad b \equiv e^{-\beta} = \exp(-h\nu/kT). \qquad (23)$$

From Eqs. (22) and (23) we have

$$\varphi_\eta(t) = \left(\frac{1-b}{1-be^{it}}\right) \varphi_\varsigma(t) \Rightarrow \varphi_\eta(t) = \langle e^{i(\xi+\varsigma)t} \rangle = \langle e^{i\xi t} \rangle \langle e^{i\varsigma t} \rangle = \varphi_\xi(t) \varphi_\varsigma(t), \qquad (24)$$

where in the first factor we recognize the chatacteristic function of the Planck-Bose distribution [10]. This way we have found a discrete random variable $\xi$ whose distribution can be easily calculated on the basis of Eq. (24):

$$\varphi_\xi(t) = \langle e^{i\xi \cdot t} \rangle = \left(\frac{1-b}{1-be^{it}}\right) = (1-b) \sum_{n=0}^\infty b^n e^{in \cdot t}, \qquad (25)$$

that is

$$f_\xi(n) \equiv p_n \equiv P(\xi = n) = (1-b)b^n = \frac{1}{1+\bar{n}}\left(\frac{\bar{n}}{1+\bar{n}}\right)^n, \quad \bar{\xi} = \bar{n} = \frac{1}{e^{h\nu/kT} - 1}, \quad b = e^{-h\nu/kT}. \qquad (26)$$

In Eq. (26) $\bar{n}$ denotes the mean photon occupation number which we have already *formally* introduced in the last equation of Eq. (20). According to Eqs.(13), (24) and (26) the total scaled energy $\eta$ of a chaotic mode of frequency $\nu$ can be split into the sum $\eta = \xi + \varsigma = [\eta] + \{\eta\}$, i. e.

$$E = h\nu\eta = h\nu(\xi + \varsigma) = h\nu([E/h\nu] + \{E/h\nu\}), \qquad (27)$$

where the discrete random variable $\xi = [\eta]$ follows the Planck-Bose distribution given in Eq. (26), and the dark part $\varsigma = \{\eta\}$ follows the (undivisible) distribution of finite support given by Eq. (14). Henceforth we shall call $\xi$ *Planck variable*.

According to Eq. (26) the average energy of the Planck-Bose part equals $E_\xi = h\nu\overline{\xi} = h\nu\overline{n}$, from which Planck' law of black-body radiation follows:

$$E_\xi = h\nu\overline{\xi} = E_\eta - E_\varsigma = h\nu\overline{n} \Rightarrow u(\nu,T) = Z_\nu E_\xi = Z_\nu h\nu\overline{n} = \frac{8\pi\nu^2}{c^3} \cdot \frac{h\nu}{e^{h\nu/kT} - 1}. \tag{28}$$

The entropy $S_\xi$ of the Planck-Bose distribution can be calculated with the help of Boltzmann's expression by taking Eq. (26) into account,

$$S_\xi = -k\sum_{n=0}^{\infty} p_n \log p_n = k[(1+\overline{n})\log(1+\overline{n}) - \overline{n}\log\overline{n}]. \tag{29}$$

It can be checked that $\partial S_\xi / \partial \overline{\xi} = k\beta = \varepsilon_0/T$, i. e., with Eq. (21) we have

$$\partial S_\varsigma / \partial E_\varsigma = \partial S_\eta / \partial E_\eta = \partial S_\xi / \partial E_\xi = 1/T, \tag{30}$$

which means that the Planck-Bose part is in thermal equilibrium with the dark part. The fluctuation formula for a system of oscillators was derived by Laue by using the Planck-Bose distribution Eq. (4). By a simple calculation we obtain

$$\overline{n^2} = \overline{n} + 2\overline{n}^2, \quad \text{hence} \quad \Delta n^2 = \overline{n^2} - \overline{n}^2 = \overline{n} + \overline{n}^2,$$

$$\Delta E_\nu^2 = M_\nu (h\nu)^2 \Delta n^2 = h\nu\overline{E}_\nu + \overline{E}_\nu^2 / M_\nu, \text{ where } \overline{E}_\nu = M_\nu h\nu \cdot \overline{n} \tag{11}$$

If we identify $M_\nu$ with the degrees of freedom of the radiation field (with the number of modes) then the physical content of Eq. (11) is the same as that of Einstein's formula.

At the beginning of this section we saw that the Gauss variable $\eta$ is infinitely divisible. Since one of the components of $\eta$, namely the dark variable $\varsigma$ is undecomposable (in an other word, *irreducibile*) because it has a finite support, the other component $\xi$, the Planck variable has to be infinitely divisible. In the followings we are going to study this question.

## 5. The infinite divisibility of the Planck variable

In the present section we prove that the integer part $\xi = [\eta]$ of the scaled energy of the chaotic field, the Planck variable can be decomposed into an infinite sum of binary random variables, which correspond to "fermion photo-molecules" (we will simply call them "binary photons") containing $2^s = 1, 2, 4, 8, ...$ single photon energies, where $s = 0, 1, 2, 3, ...$ . We have received a hint for this decomposition from the books by Székely [24] and Lukács [25]. At this point we note that a detailed general analysis can be found on the infinitely divisible random varables in the book by Gnedenko and Kolmogorov [26].

With the help of the algebraic identity $(1-z)(1+z)(1+z^2)\cdot...\cdot(1+z^{2^s}) = 1 - z^{2^{s+1}}$ we have the following absolutely convergent infinite product representation of $1/(1-z)$

$$\frac{1}{1-z} = \prod_{s=0}^{\infty}(1+z^{2^s}) \quad (|z|<1) . \tag{31}$$

By using Eq. (31), the characteristic function of the Planck-Bose distribution of the random variable $\xi$, Eq. (25), can be expanded into the infinite product of characteristic functions

$$\frac{1-b}{1-be^{it}} = \lim_{q\to\infty} \frac{1-b^{2^{q+1}}}{1-b^{2^{q+1}}e^{2^{q+1}it}} \prod_{s=0}^{q} \frac{1+b^{2^s}e^{2^s it}}{1+b^{2^s}} = \prod_{s=0}^{\infty}\frac{1+b^{2^s}e^{2^s it}}{1+b^{2^s}}, \quad (0<b<1), \tag{32}$$

that is

$$\frac{1-b}{1-be^{it}} = \prod_{s=0}^{\infty}\frac{1+b^{2^s}e^{2^s it}}{1+b^{2^s}} = \frac{1+be^{it}}{1+b}\cdot\frac{1+b^2 e^{2it}}{1+b^2}\cdot\frac{1+b^4 e^{4it}}{1+b^4}\cdot..., \quad b = \exp(-h\nu/kT) . \tag{33}$$

From Eqs. (25) and (33) we obtain

$$\varphi_\xi(t) = \varphi_{u_0}(t)\varphi_{u_1}(t)\cdots\varphi_{u_s}(t)\cdots, \quad \text{or} \quad \varphi_\xi(t) = \left\langle e^{i\xi \cdot t} \right\rangle = \left\langle e^{i(u_0+u_1+\ldots u_s+\ldots)\cdot t} \right\rangle, \tag{34}$$

where

$$\varphi_{u_s}(t) = \left\langle e^{iu_s t} \right\rangle = \frac{1+b^n e^{in\cdot t}}{1+b^n}, \quad (b=\exp(-h\nu/kT)), \quad (n \equiv 2^s). \tag{35}$$

Hence, by a proper choice of the sample (event) space, the random variable $\xi$ can be decomposed into an infinite sum of completely independent variables $\{u_s ; s = 0, 1, 2,...\}$

$$\xi = u_0 + u_1 + u_2 + \ldots + u_s + \ldots, \tag{36}$$

which have the binary distributions

$$p_0(s) = P(u_s = 0) = \frac{1}{1+b^{2^s}}, \quad p_1(s) = P(u_s = 2^s) = \frac{b^{2^s}}{1+b^{2^s}} \quad (s=0, 1, 2,...). \tag{37}$$

One can easily check that the characteristic function of these variables are really the factors of the product Eq. (34) given by Eq. (35). The probabilities in Eq. (37) can be written out in detail as

$$P(u_s = 0) \equiv P(\overline{A}_s; u_s(\overline{A}_s) = 0), \quad P(u_s = 2^s) \equiv P(A_s; u_s(A_s) = 2^s), \tag{38}$$

where $A_s$ denotes the event that the *s*-th binary component is occupied by one excitation with energy $2^s h\nu$, and $\overline{A}_s = I - A_s$ denotes the complementary event, i.e. that event when the *s*-th binary component is not occupied. In short, Eq. (38) can be expressed as

$$P(\overline{A}_s) = \frac{1}{1+b^{2^s}}, \quad P(A_s) = \frac{b^{2^s}}{1+b^{2^s}}. \tag{39}$$

In this description the excitation events form a $\sigma$- algebra in Kolmogorov sense (see for instance the books by Rényi [27] and Feller [28]. If the original bosonic random variable $\xi$ takes the value $\xi(B_n) = n$, then this circumstance corresponds to the event $B_n$, that is, the particular mode of the black-body radiation is excited exactly to the *n*-th level of energy $nh\nu$.

Since any integer number *n* can be expanded into a sum of powers of *2* (this is called the dyadic expansion), the bosonic excitations can be uniquely expressed in terms of the binary excitations. For example, if exactly $9 = 1 \cdot 2^0 + 0 \cdot 2^1 + 0 \cdot 2^2 + 1 \cdot 2^3$ photons are excited in the mode, then this event is expressed as the following product

$$B_9 = A_0 \overline{A}_1 \overline{A}_2 A_3 \overline{A}_4 \overline{A}_5 ... \overline{A}_s ... \ . \tag{40}$$

By using Eq. (39), the probability of the event $B_9$ equals

$$P(B_9) = P(A_0)P(\overline{A}_1)P(\overline{A}_2)P(A_3)\prod_{s=4}^{\infty} P(\overline{A}_s) = \frac{P(A_0)P(A_3)}{P(\overline{A}_0)P(\overline{A}_3)} \prod_{s=0}^{\infty} P(\overline{A}_s),$$
$$= b^{2^0} b^{2^3} (1-b) = (1-b)b^9 \tag{41}$$

in complete accord with the original Planck-Bose distribution, Eq. (26). As another example, let us consider the event $(A_0 + A_3)\overline{A}_1 \overline{A}_2 \overline{A}_4 \overline{A}_5 ... \overline{A}_s ...$, which means the non-exlusive alternative that *either* the 0-th *or* the 3-rd multiplet components or *both of them* are excited, and all the other components (*s = 1, 2, 4, 5, …*) are unoccupied at the same time. By using the rules of the usual Boole algebra of events we obtain

$$\begin{aligned}(A_0 + A_3)\overline{A}_1 \overline{A}_2 \overline{A}_4 \overline{A}_5 ... \overline{A}_s ... &= A_0 \overline{A}_1 \overline{A}_2 A_3 \overline{A}_4 \overline{A}_5 ... \overline{A}_s ... + A_0 \overline{A}_1 \overline{A}_2 \overline{A}_3 \overline{A}_4 \overline{A}_5 ... \overline{A}_s ... \\ &+ A_0 \overline{A}_1 \overline{A}_2 A_3 \overline{A}_4 \overline{A}_5 ... \overline{A}_s ... + \overline{A}_0 \overline{A}_1 \overline{A}_2 A_3 \overline{A}_4 \overline{A}_5 ... \overline{A}_s ... \\ &= A_0 \overline{A}_1 \overline{A}_2 A_3 \overline{A}_4 \overline{A}_5 ... \overline{A}_s ... + A_0 \overline{A}_1 \overline{A}_2 \overline{A}_3 \overline{A}_4 \overline{A}_5 ... \overline{A}_s ... + \overline{A}_0 \overline{A}_1 \overline{A}_2 A_3 \overline{A}_4 \overline{A}_5 ... \overline{A}_s ... \\ &= B_9 + B_1 + B_8 \end{aligned} \tag{42}$$

From Eq. (39) the corresponding probability can simply be calculated

$$P((A_0 + A_3)\overline{A}_1 \overline{A}_2 \overline{A}_4 \overline{A}_5 ... \overline{A}_s ...) = (1-b)(b + b^8 + b^9) = P(B_1 + B_8 + B_9). \tag{43}$$

The result, Eq. (43) can also be obtained directly from the original Planck-Bose distribution, Eq. (26), by assuming that the different photon excitations correspond to completely independent events in the present case, i. e.

$$P(B_1 + B_8 + B_9) = P(B_1) + P(B_8) + P(B_9) = (1-b)(b + b^8 + b^9). \tag{44}$$

In the standard description on the basis of quantized modes, Eq. (44) can be expressed with the help of the photon number projectors as

$$P(B_1 + B_8 + B_9) = Tr[\rho(\Pi_1 + \Pi_8 + \Pi_9)], \qquad \Pi_n \equiv |n\rangle\langle n|, \tag{45}$$

where $\rho$ denotes the density operator of the thermal mode:

$$\rho = \sum_{n=0}^{\infty} |n\rangle(1-b)b^n \langle n| = \sum_{n=0}^{\infty} |n\rangle \frac{\bar{n}^n}{(1+\bar{n})^{n+1}} \langle n|, \qquad b = \exp(-h\nu/kT). \tag{46}$$

From Eq. (45) it is clear that the mutually orthogonal projectors $\Pi_n$ represent the mutually independent events $B_n$. We note that for the compactness of our formulae, we will henceforth sometime use the notation $n \equiv 2^s$. The expectation value of $u_s$ is easily determined from Eq. (37)

$$\bar{u}_s = p_0(s) \cdot 0 + p_1(s) \cdot 2^s = \frac{nb^n}{1+b^n} = b\frac{\partial}{\partial b}\log(1+b^n), \qquad (n \equiv 2^s). \tag{47}$$

According to Eq. (36), the sum of these expectation values, of course, is equal to $\bar{\xi}$ ( given in Eq. (3.5) ), which can be shown by direct calculation by using Eq. (47),

$$\sum_{s=0}^{\infty} \bar{u}_s = b\frac{\partial}{\partial b}\sum_{s=0}^{\infty}\log(1+b^n) = b\frac{\partial}{\partial b}\left\{\log\left[\prod_{s=0}^{\infty}(1+b^n)\right]\right\} = b\frac{\partial}{\partial b}\log\left(\frac{1}{1-b}\right) = \frac{1}{b^{-1}-1} = \bar{\xi}. \tag{48}$$

The expectation value of the energy of the *s*-th fermion multiplet is

$$E_s = h\nu\bar{u}_s = 2^s h\nu \frac{1}{\exp(2^s h\nu/kT)+1} = 2^s h\nu\bar{n}_s, \qquad \bar{n}_s = \frac{1}{\exp(2^s h\nu/kT)+1}. \tag{49}$$

According to Eq. (49), the fermion multiplet gases have zero chemical potential. Their entropies are all binary entropies, which read

$$\begin{aligned}S_s &= -k\{p_0(s)\log[p_0(s)] + p_1(s)\log[p_1(s)]\} \\ &= -k\{[1-(E_s/2^s h\nu)]\log[1-(E_s/2^s h\nu)] + (E_s/2^s h\nu)\log(E_s/2^s h\nu)\}, \\ &= -k[(1-\bar{n}_s)\log(1-\bar{n}_s) + \bar{n}_s\log\bar{n}_s]\end{aligned} \tag{50}$$

where the probabilities have been given in Eq. (37), and $E_s$ and $\bar{n}_s$ are defined in Eq. (49). It is useful to have still another form of the entropies, which can be easily summed up,

$$S_s = -k\left[\log\left(\frac{1}{1+b^n}\right) + \frac{nb^n}{1+b^n}\log b\right], \quad (n \equiv 2^s), \quad (b = \exp(-h\nu/kT)). \tag{51}$$

From Eqs. (50) and (51) it can be proved that *the thermodynamic relation*

$$dS_s / dE_s = 1/T \tag{52}$$

*is satisfied as an identity for all s-values.* The sum of the entropies of the fermion multiplets is exactly equal to the entropy of the Planck-Bose distribution.

$$\sum_{s=0}^{\infty} S_s = k\log\left[\prod_{s=0}^{\infty}(1+b^n)\right] - k\left\{b\frac{\partial}{\partial b}\log\left[\prod_{s=0}^{\infty}(1+b^n)\right]\right\}\log b,$$
$$= k\log\left(\frac{1}{1-b}\right) - k\left(\frac{b}{1-b}\right)\log b = S \tag{53}$$

i. e.

$$S = \sum_{s=0}^{\infty} S_s. \tag{54}$$

Eq. (54) means that the fermion components are thermodynamically independent.

The second expression in Eq. (52) for the entropy $S_s(E_s)$ can also be derived on the basis of Planck's general definition of entropy [19] in the following way. Let us distribute the $P$ fermion excitations of energy $\varepsilon_s = 2^s h\nu$ among $M = V(8\pi\nu^2 d\nu/c^3)$ modes in the volume $V$ and in the frequency interval $(\nu, \nu + d\nu)$. This is just the number of combinations when we distribute $P$ undistingvishable object into $M$ places, namely $M!/P!(M-P)!$. The total energy is assumed to consist of $P$ elementary excitations, that is, $ME_s = P\varepsilon_s$. By using Stirling's formula $M! \approx (M/e)^M$, the entropy of one mode can be expressed as

$$S_s = \frac{1}{M} k \log\left[\binom{M}{P}\right] = -k\left[\left(\frac{P}{M}\right)\log\left(\frac{P}{M}\right) + \left(1-\frac{P}{M}\right)\log\left(1-\frac{P}{M}\right)\right], \tag{55}$$
$$= -k[(1 - E_s/\varepsilon_s)\log(1 - E_s/\varepsilon_s) + (E_s/\varepsilon_s)\log(E_s/\varepsilon_s)]$$

which coincides with the second expression of Eq. (50). From (55) and from the thermodynamic relation $dS_s/dE_s = 1/T$ we obtain, of course the averaged energy.

When we calculate the complete fluctuation of the original (bosonic) field, this turns out to be an infinite sum of fermion-type fluctuations of the binary photons,

$$\Delta\xi^2 = \bar{n} + \bar{n}^2 = \sum_{s=0}^{\infty} \Delta\bar{u}_s^2 = \sum_{s=0}^{\infty}(2^s\bar{u}_s - \bar{u}_s^2). \tag{55a}$$

The Fermi distribution Eq. (49) can also be obtained by using the reaction kinetic considerations due to Ornstein and Kramers [18]. Let us consider four modes $M_1', M_2'$ and $M_1'', M_2''$ among the set of $M$ modes in the frequency interval $(\nu, \nu+d\nu)$. Assume that the first two modes are excited by the fermion photo-multiplets $s_1', s_2'$, and – due to the (at the moment not specified) interaction with a small black body ("Planck's Kohlenstäubchen", a small carbon particle) – they give their energy through an other pair of excitations $s_1'', s_2''$ occupying the other two modes $M_1'', M_2''$. Because of the dynamical equlibrium, the reversed process can also take part with the same probability per unit time. The rate of these processes

$$R(\rightarrow) = w(\rightarrow)\bar{n}_{s_1'} \cdot \bar{n}_{s_2'} \cdot (1 - \bar{n}_{s_1''}) \cdot (1 - \bar{n}_{s_2''}), \tag{56}$$

$$R(\leftarrow) = w(\leftarrow)\bar{n}_{s_1''} \cdot \bar{n}_{s_2''} \cdot (1 - \bar{n}_{s_1'}) \cdot (1 - \bar{n}_{s_2'}), \tag{57}$$

where, of course, we require the energy conservation $\varepsilon_{s_1'} + \varepsilon_{s_2'} = \varepsilon_{s_1''} + \varepsilon_{s_2''}$ to be satisfied. In dynamical equilibrium the two rates Eq. (56) and (57) must be equal, moreover we assume that the transition probabilities of the "direct" and the reversed processes are equal,

$$R(\rightarrow) = R(\leftarrow), \ w(\rightarrow) = w(\leftarrow). \tag{59}$$

Then, by introducing the quantities

$$q_s \equiv \frac{\bar{n}_s}{1-\bar{n}_s} , \tag{60}$$

from Eqs. (59-60) we obtain

$$q_{s'_1} q_{s'_2} = q_{s''_1} q_{s''_2} . \tag{61}$$

Equation (61) can be satisfied for all $s'_1, s'_2, s''_1, s''_2$ – satisfying the subsidiary condition $\varepsilon_{s'_1} + \varepsilon_{s'_2} = \varepsilon_{s''_1} + \varepsilon_{s''_2}$ – when we choose the only reasonable solution characterized by the two parameters $\alpha$ and $\beta$. With $\alpha = 0$ and $\beta = 1/kT$, Eq. (49) can be recovered,

$$q_s = e^{-\alpha - \beta \varepsilon_s} \rightarrow \bar{n}_s = \frac{1}{e^{\varepsilon_s/kT} + 1} . \tag{62}$$

## 6. Summary

In the present paper we have first given in the introduction a short summary on the early theories on the black-body radiation and quantum statistics, and presented our motivation to study this subject today. We have quoted the original literature on purpose in order to provide the interested reader with the references where original thougths on the concept of quanta first appeared. In Section 2 we have introduced our basic notations and the main approach of our paper based on classical probability theory. For completeness we have outlined the derivation of the chaotic (Gauss) distribution from the central limit theorem for one mode of the black-body radiation and proved the energy equipartition theorem and the Rayleigh-Jeans law. We have also derived the wave-like fluctuation term of the energy of the chaotic field in a sub-volume of a cavity. In Section 3 the construction of the fractional part of the Gauss variable has been carried out. We call the random variable so obtained, the "dark variable" which is already an undecomposable (irreducible) random variable of finite support. We have seen that the average energy of the "dark part" approaches from below the zero-point energy as we let the absolut temperature going to infinity. In Section 4 we have derived the well-known Planck-Bose distribution of the Planck variable which has been proved to be the integer part of the Gauss variable. Here we have also proved that the dark part and the Planck part are thermodynamically independent, since their entropy simply add to yield the entropy of the Gauss variable belonging to one spectral component of the chaotic field. In Section 5 it has been shown that the Planck-Bose distribution is infinitely divisible, and the Planck variable can be decomposed into an infinite sum of independent binary random variables representing the "binary photons". These binary photons follow the Fermi statistics, which has been derive in two other different ways, too. It has been proved that each spectral components of the black-body radiation can be viewed as a mixture of thermodynamically independent fermion gases consisting of binary photons. In this way we have presented a unique irreducible decomposition of the Gauss variable into a sum of one dark variable and the infinite assembly of binary variables. In this section we have also considered a simple example to illustrate the dyadic expansion of ordinary photon excitations in terms of the binary photon excitations. We note that the above analysis may give a hint to consider quantum events in some cases as "ordinary events" in the Kolmogorov sense. Of course, we have not been able to discuss interference of events in terms of classical random variables. We have merely applied the $\sigma$-algebra for the events appearing at the act of measurement (when the projection onto a photon number state happens). Hence our analysis is limited to describe incoherent excitations which is, on the other hand, just relevant in the case of black-body radiation. At this point let us mention that there has been an extensive study developing on the *quantum* Gauss distributions which has been rececently discussed in details for instance by Wolf et al. [30]. These Gauss distributions are in fact Wigner functions with a Hilbert space backgroun of quantum states. Nevertheless, as we have seen in Section 5 , the quantum excitations discussed by us can be represented by a classical $\sigma$-algebra yielding the same results as the application of the usual quantum projectors.

**Acknowledgement**

This work has been supported by the Hungarian National Scientific Research Foundation (OTKA, grant number T048324). I thank Prof. J. Javanainen of University of Connecticut for several useful discussions on the new derivation of black-body spectrum. I also thank Prof. G. J. Székely of Bowling Green State University for drawing my attention to the mathematical literature on the irreducible decomposition of the discrete exponential distribution.

Sándor Varró : Irreducible Decomposition of Gaussian Distributions 27[8] A. Einstein, Zum gegenwärtigen Stand des Strahlungsproblems, Phys. Zeitschr. **10** (1909) 185-193.

[9] A. Einstein, Zur Quantentheorie der Strahlung, Phys. Zeitschr. **18** (1917) 121-128.

[10] P. Debye, Der Wahrscheinlichkeitsbegriff in der Theorie der Strahlung, Ann. der Phys. **33** (1910) 1427-1434.

[11] M. Wolfke, Einsteinsche lichtquanten und räumliche Struktur der Strahlung, Phys. Zeitschr. **22** (1921) 375-379.

[12] W. Bothe, Die räumliche Energieverteilung in der Hohlraumstrahlung, Zeitschr. für Phys. **20** (1923) 145-152.

[13] L. Natanson, Über die statistische Theorie der Strahlung, Phys. Zeitschr. **11** (1911) 659-666. Translated from the original : On the Statistical Theory of Radiation, Bulletin de l'Académie des Sciences de Cracovie (A) (1911) 134-148.

[14] S. N. Bose, Plancks Gesetz und Lichtquantenhypothese, Zeitschr. für Phys. **26** (1924) 178-181.

[15] A. Einstein, Quantentheorie des einatomigen idealen Gases, Sitzungsberichte der Preuss. Akad. Wiss. XXII (1924) 261-267.

[16] A. Einstein, Quantentheorie des einatomigen idealen Gases. Zweite Abhandlung, Sitzungsberichte der Preuss. Akad. Wiss. XXIII (1925) 3-14.

[17] E. Fermi, Zur Quantelung des idealen einatomigen Gases, *Zeitschr. für Phys.* **36** (1926) 902-912.

[18] L. S. Ornstein und H. A. Kramers, Zur kinetischen Herleitung des Fermischen Verteilungsgesetzes, Zeitschr. für Phys. **42** (1927) 481-486.

[19] M. Planck, *Theorie der Wärmestrahlung* (Johann Ambrosius Barth-Verlag, Leipzig, 1906/66 ), see also M. Planck, *The theory of heat radiation* (Dover Publications, Inc., New York, 1959), translation of the second edition of Planck's book by M. Masius from 1914.